\documentclass[english]{article}

\usepackage{array}
\usepackage{amsmath}
\usepackage{bm}
\usepackage{cancel}
\usepackage{graphicx}

\usepackage{color}
\definecolor{headingcolour}{RGB}{50, 150, 190}

\usepackage{lipsum}

\usepackage{babel}

\usepackage[authoryear]{natbib}

\usepackage{setspace}

\usepackage{quattrocento}

\usepackage{sectsty}
\allsectionsfont{\fontfamily{phv}\selectfont}

\usepackage[T1]{fontenc} 

\usepackage[utf8]{inputenc}

\usepackage[margin=1in]{geometry}
\usepackage{fancyhdr}
\usepackage{caption}
\DeclareCaptionFormat{label}{\textbf{#1 #2} #3\hrulefill}
\usepackage[compact]{titlesec}

\usepackage{graphicx}

\usepackage{pifont}

\usepackage{enumitem}
\setlist{nosep}

\renewenvironment{abstract}{%
	\hspace{0.025\linewidth}\begin{minipage}{0.95\textwidth}
		\rule{\textwidth}{1pt}\small\selectfont}
	{\vspace{-0.5em}\par\noindent\rule{\textwidth}{1pt}\end{minipage}\vspace{1em}}
\makeatletter
\renewcommand{\maketitle}{\bgroup\setlength{\parindent}{0pt}
	\thispagestyle{empty}
	\begin{flushleft}
		{\bf \fontfamily{phv}\selectfont \LARGE \@title}
		
		\bf \fontfamily{phv}\selectfont \@author
	\end{flushleft}\egroup
}
\makeatother
\usepackage[figuresright]{rotating}
\usepackage{siunitx}
\usepackage[%
    colorlinks=true,
    pdfborder={0 0 0},
    allcolors =cyan
]{hyperref}
\usepackage{dingbat}

\let\ts\textsubscript

\usepackage[flushleft]{threeparttable}

\usepackage{lscape} 

\usepackage{lineno}
\usepackage{authblk}

\newcommand*\patchAmsMathEnvironmentForLineno[1]{%
  \expandafter\let\csname old#1\expandafter\endcsname\csname #1\endcsname
  \expandafter\let\csname oldend#1\expandafter\endcsname\csname end#1\endcsname
  \renewenvironment{#1}%
     {\linenomath\csname old#1\endcsname}%
     {\csname oldend#1\endcsname\endlinenomath}}%
\newcommand*\patchBothAmsMathEnvironmentsForLineno[1]{%
  \patchAmsMathEnvironmentForLineno{#1}%
  \patchAmsMathEnvironmentForLineno{#1*}}%
\AtBeginDocument{%
\patchBothAmsMathEnvironmentsForLineno{equation}%
\patchBothAmsMathEnvironmentsForLineno{align}%
\patchBothAmsMathEnvironmentsForLineno{flalign}%
\patchBothAmsMathEnvironmentsForLineno{alignat}%
\patchBothAmsMathEnvironmentsForLineno{gather}%
\patchBothAmsMathEnvironmentsForLineno{multline}%
}

\title{Impacts of permeability heterogeneity and background flow on supercritical CO\textsubscript{2} dissolution in the deep subsurface}

\author[Hansen et al.,]{S.~K.~Hansen$^{1,*}$, Y.~Tao$^{1,\dag}$, and  S.~Karra$^2$\\
{\small $^1$Zuckerberg Institute for Water Research, Ben-Gurion University of the Negev, Sede Boqer Campus, Israel 8499000.} \\
{\small $^2$Environmental Molecular Sciences Laboratory, Pacific Northwest National Laboratory, Richland, WA, USA 99352.} \\
{\small $^*$Corresponding author:~Scott K. Hansen, Email:~\texttt{skh@bgu.ac.il}}\\
{\small $^\dag$Now at~Nelson Institute for Environmental Studies, University of Wisconsin--Madison}}

\begin{document}
\doublespacing
\maketitle
\onehalfspacing

\begin{abstract} 
	Motivated by CO\ts{2} capture and sequestration (CCS) design considerations, we consider the coupled effects of permeability heterogeneity and background flow on the dissolution of a supercritical CO\ts{2} lens into an underlying deep, confined aquifer. We present the results of a large-scale Monte Carlo simulation study examining the interaction of background flow rate and three parameters describing multi-Gaussian log-permeability fields: mean, variance, and correlation length. Hundreds of high-resolution simulations were performed using the PFLOTRAN finite volume software to model CO\ts{2} dissolution in a kilometer-scale aquifer over 1000 y. Predictive dimensionless scaling relationships relating CO\ts{2} dissolution rate to heterogeneity statistics, Rayleigh (Ra) and P\'{e}clet (Pe) numbers were developed for both gravitationally dominated free convection to background flow-dominated forced convection regimes. An empirical criterion, $\rm Pe\ = Ra^{3/4}$, was discovered for regime transition. All simulations converged quickly to a quasi-steady, approximately linear dissolution rate. However, this rate displayed profound variability between permeability field realizations sharing the same heterogeneity statistics, even under mild permeability heterogeneity. In general, increased heterogeneity was associated with a lower mean and higher variance of dissolution rate, undesirable from a CCS design perspective. The relationship between dissolution rate and background flow was found to be complex and nonlinear. Dimensionless scaling relationships were uncovered for a number of special cases. Results call into question the validity of the Boussinesq approximation in the context of modest-to-high background flow rates and the general applicability of numerical simulations without background flow.
\end{abstract}


\section{Introduction} 
	The behavior of CO\textsubscript{2} injected into confined saline aquifers has become a topic of intense research interest in the past two decades, motivated in part by a desire to design carbon capture and storage (CCS) sequestration schemes that prevent CO\textsubscript{2} from entering the atmosphere. CO\textsubscript{2} is soluble in water, and the dissolution of CO\textsubscript{2} from a buoyant supercritical phase into aquifer brine is recognized as a crucial trapping mechanism. Due to gravitational instability, convective fingering of the CO\textsubscript{2}-enriched brine is expected, which may enhance uptake: increasing the interfacial area between CO\textsubscript{2}-rich and CO\textsubscript{2}-poor regions and also driving large-scale mixing.	The relevant governing equations for density-driven flow and transport are nonlinear \citep{Hidalgo2009} and so can only be fully studied by direct numerical simulation (DNS). Most commonly, DNS of supercritical CO\textsubscript{2} dissolution is performed in 2D domains with spatially homogeneous permeability and no background flow. However, recent numerical and experimental work has suggested that (i) permeability heterogeneity may have a dominant effect on fingering \citep{Jahangiri2011,Aggelopoulos2012,Taheri2012,Kong2013,Zhang2013,Song2014,Frykman2014,Salibindla2018,Singh2021,Wang2021}, that (ii) background flow may either increase trapping by enhancing transverse dispersion or inhibit it by suppressing fingering \citep[e.g.,][]{Hassanzadeh2009,Emami-Meybodi2015a,Emami-Meybodi2017a,Michel-meyer2017,Michel-Meyer2020}, and that, (iii) the effect of local-scale longitudinal dispersion has recently been identified as suppressing convection in homogeneous systems \textit{with} background flow \citep{Tsinober2022} and as enhancing it in heterogeneous systems without it \citep{Erfani2021}.
			
	While there is limited published DNS work examining these factors individually, no study we are aware of addresses their combined effect in the context of CO\textsubscript{2}. Considering the combined effect is important for three reasons. First, we have found that even mild background flow in a heterogeneous environment completely disrupts convective finger formation, suggesting limited relevance of DNS without background flow. Second, these investigations have also shown (Figure \ref{fig: transition}) uptake rate to be strongly reduced at the regime transition where background flow begins to disrupt gravitational convection (\cite{Tsinober2022} show a qualitatively similar trend). Third, we may expect a strong increase in aquifer CO\textsubscript{2} intake due to transverse macrodispersive mixing---which generally will enhance dissolution, and is a potentially much stronger effect than has been studied in quasi-homogeneous media \citep[e.g.,][]{Michel-meyer2017}. To understand the competing effects of the suppression of gravitational convection and the addition of macrodispersion due to background flow, as well as to characterize the transition between regimes, DNS that explicitly considers all factors together---permeability heterogeneity structure, local-scale dispersion, and strength of background flow---is required.

	Study of CO\textsubscript{2} behavior in the subsurface commonly focuses on vertical convective fingering and employs the simplifying assumptions of no background flow and homogeneous permeability \citep[e.g.,][]{Ennis-King2005,Riaz2006,Pau2010,Teres2012,Hidalgo2012,Soltanian2016,Macminn2018}. Another popular approach considers essentially horizontal flow in a homogeneous \citep{Tutolo2015} or layered system \citep{Vilarrasa2013, Song2014}, neglecting convective fingering altogether. Much has been learned from these simplified approaches; however, the relevance of permeability heterogeneity and background flow to CO\textsubscript{2} sequestration is becoming increasingly clear.
	
	Hydrogeological interest in unstable, potentially convective flow and transport predates the recent interest motivated by CCS schemes. In particular, the conditions under which fingering occurs in variable-density miscible flows were studied in the subsurface context by \cite{Oostrom1992} and \cite{Schincariol1994}. Similar behavior was considered in the context of immiscible displacement by \cite{chen1996wetting}, \cite{neuman1996instability}, and \cite{chen2000wetting}. In recent years, interest has intensified in these topics.
	
	A small number of numerical papers have considered the effect of background flow on CO\textsubscript{2} dissolution and uptake. Its impact on the onset of convection was studied in a homogeneous permeability field by \cite{Hassanzadeh2009}, who employed linear stability analysis. The authors found that transverse dispersion increased the time to onset of convection and affected the wavelength of convective fingers. \cite{Rapaka2011} identified a critical ratio of background flow velocity to buoyancy velocity above which convective fingering is suppressed. \cite{Emami-Meybodi2015a} developed a semi-analytic solution for CO\textsubscript{2} dissolution in this suppressed-fingering regime. \cite{Emami-Meybodi2017a} presented a stability analysis supplemented by numerical simulation to characterize finger formation under weaker background flows. All of these analyses, we note, neglected permeability field heterogeneity. Most recently, \cite{Tsinober2022} presented a numerical analysis considering the combined effect of background flow and local-scale dispersion on uptake rate. All of these analyses were performed in homogeneous permeability fields. 
	
	A range of numerical work directly considers permeability heterogeneity. In an early study that predates the flourishing of the CCS literature, \cite{Schincariol1998}  explicitly considered the evolution of dense solute plumes in heterogeneous media with background flow. Strong variability in the convective mixing strength, even between different realizations with the same heterogeneity statistics, was noted, and the classical stability criteria were inappropriate. \cite{Simmons2001} considered the impact variability and correlation structure of the permeability field in the absence of background flow and found convective mixing strongly dependent on both statistics. \cite{Prasad2003} extended these studies, again in the context of no background flow, finding convective behavior to be strongly affected by permeability field correlation length (due to the suppressive effect of horizontal low-permeability features). They also observed that increased permeability variance had opposite effects on convection at early vs. late time. Turning explicitly to the context of DNS of CO\textsubscript{2}, \cite{Jahangiri2011} considered the effects of permeability heterogeneity on horizontal macrodispersive spreading (as opposed to convective fingering) of a dissolved CO\textsubscript{2} plume. \cite{Tian2016} considered a similar scenario, using a full reactive flow and transport simulator to analyze horizontal fingering due to permeability heterogeneity. Limited modeling has considered the effects of permeability heterogeneity on vertical, convection-driven fingering of CO\textsubscript{2}. \cite{Taheri2012} performed DNS without background flow on several simplified heterogeneous scenarios (e.g., homogeneous aquifers with thin barriers, uniform vertical or layers of different permeability), finding a strong impact of permeability heterogeneity on convection. \cite{Frykman2014} performed a numerical study of convection in a broadly homogeneous system with thin, horizontal, low-permeability lenses, finding that these greatly disrupted the convective mixing mechanism. \cite{Kong2013} presented a sophisticated numerical study that considered convective fingering in a 2D, isotropic aquifer with a multi-Gaussian correlation structure and no background flow (i.e., initially hydrostatic conditions), relating permeability log-variance and correlation length to convective fingering behavior. More recently, \cite{Soltanian2016} related vertical convective finger spreading to permeability structure in bimodal heterogeneous media without background flow. \cite{Singh2021} characterized the dissolution of a realistic 2D lens whose position in a heterogeneous permeability field was determined by the solution of two-phase flow equations during injection. The impact of permeability on subsequent dissolution and convection (absent background flow) was studied, with the authors reporting that high heterogeneity reduced the uptake relative to homogeneous conditions. To our knowledge, there are no DNS studies of permeability heterogeneity featuring background flow or employing 3D fields.
	
	In the context of CO\textsubscript{2} sequestration, physical experiments are less common than numerical studies, although a few notable ones have been published. \cite{Aggelopoulos2012} performed vertical CO\textsubscript{2} convection experiments in various macroscopically-homogeneous sand and bead packs, finding that even changing the microscale heterogeneity owing to the packing media significantly impacted CO\textsubscript{2} dissolution patterns. \cite{Salibindla2018} recently experimentally considered the impact of a low-permeability feature in a CO\textsubscript{2}-analog experimental setup, confirming numerical predictions that this may be the dominant control on convective mixing. \cite{Michel-meyer2017} performed a bench-scale horizontal column experiment with an analog fluid in a bead pack, which showed background flow suppressing convective fingering. A follow-up paper, \cite{Michel-Meyer2020} experimentally considered the impact of local-scale dispersion and quantified the total dissolution of the analog fluid, again in a homogeneous domain. Recently, \cite{Wang2021, Wang2021a} measured convective finger propagation by MRI without flow in zonated heterogeneous media, identifying a profound influence of heterogeneity. To our knowledge, no bench-scale experiments have been performed considering macroscopic heterogeneity with background flow.

	The principal aim of this study completion of a suite of simulation/regression studies that examine the effects of spatially heterogeneous aquifer permeability and background flow on the trapping of CO\textsubscript{2}, which has been injected as part of a CCS scheme. We describe a large-scale parametric study using random field generators, the state-of-the-art reactive flow and transport simulator \textsc{PFLOTRAN} \citep{Lichtner2015}, and high-performance computing to examine the effects of permeability heterogeneity, permeability correlation structure, and strength of background flow on CO\textsubscript{2} fingering dynamics and the short- and long-term rates of CO\textsubscript{2} dissolution.

	While the literature has shown them to be individually important, there is no experimental or numerical research that we are aware of which simultaneously considers convective fingering, background flow, and permeability heterogeneity (especially vertical anisotropy and correlated, low-permeability features). We aim to remedy this shortcoming and to relate convective fingering patterns and (short and long-term) CO\textsubscript{2} trapping rates to permeability field heterogeneity, correlation structure, and mean background flux in the aquifer, and ultimately to determine conditions under which trapping will be optimized.

	\begin{figure} 
		\centering 
		\includegraphics[width=\linewidth]{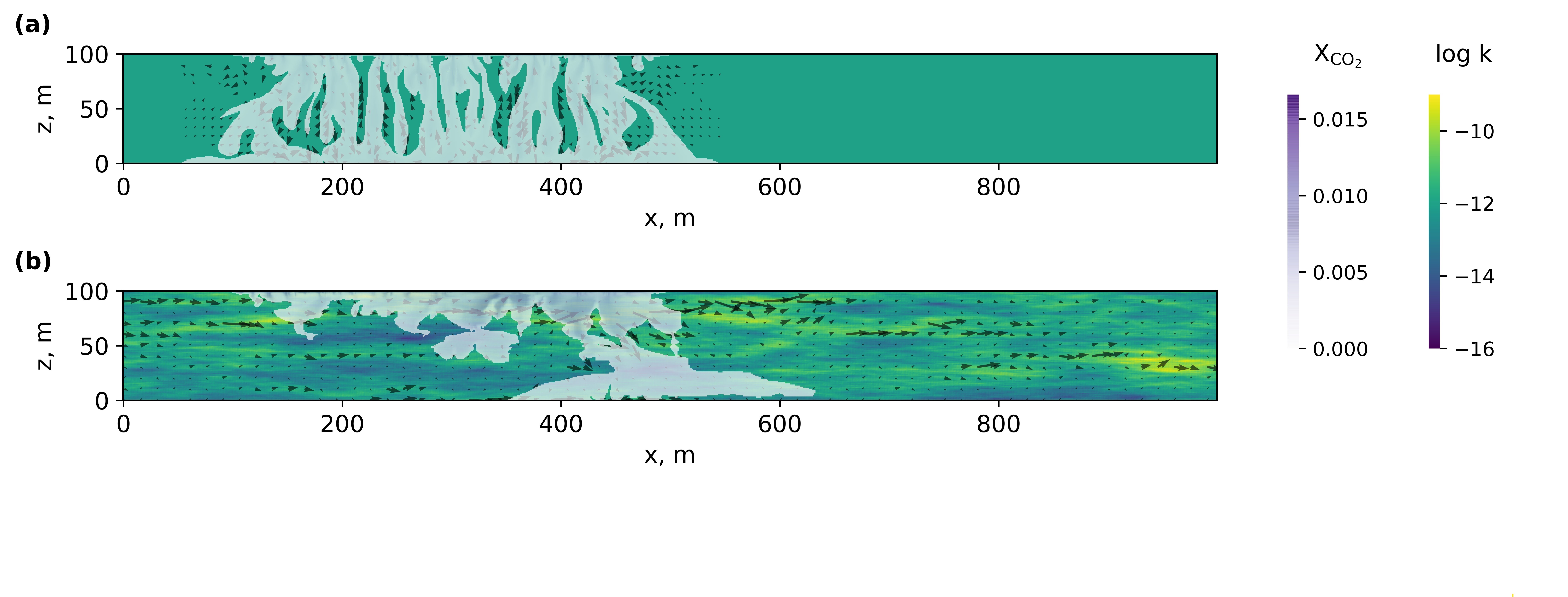} 
		\caption{CO\textsubscript{2} fingering 200 y after injection into geometric mean $10^{-12} \mathrm{m^2}$ permeability fields: (a) homogeneous, without background flow, and (b) mildly heterogeneous with approximate $10^{-2}\ \mathrm{m y^{-1}}$ background Darcy flux. Cell-center velocities are indicated by quiver arrows where they areuations significantly non-zero.} 
	\end{figure}

\section{Theory}

\subsection{Governing equations}

	Under constant-temperature, single-phase conditions---those relevant to our study---the energy balance equation may be greatly simplified, and only the balance of mass for water and CO\textsubscript{2} need to be considered. The relevant equations can be written as follows:
	\begin{eqnarray}
		\frac{\partial}{\partial t}\left(\theta c\right) + \bm{\nabla}\cdot\left(\bm{q}c - \theta D \bm{\nabla} c\right) &=& 0,\label{eq: PFLOTRAN mass balance}\\
		\bm{q} &=& \frac{k}{\mu}\bm{\nabla}\left(P+\rho g {z}\right), \label{eq: PFLOTRAN flux}\\
		\bm{\nabla}\cdot\bm{q} &=& 0,
	\end{eqnarray}
	where $\theta$ denotes porosity, $P$ pressure, $c$ concentration of CO\textsubscript{2} in water, $\mu$  viscosity of water, $k$  permeability of the medium, $\bm{q}$  Darcy flux, $g$   acceleration due to gravity, $z$  vertical coordinate, $\rho$  mass density of water and $D$ diffusion coefficient.

\begin{figure}[h]
		\includegraphics[width=\linewidth]{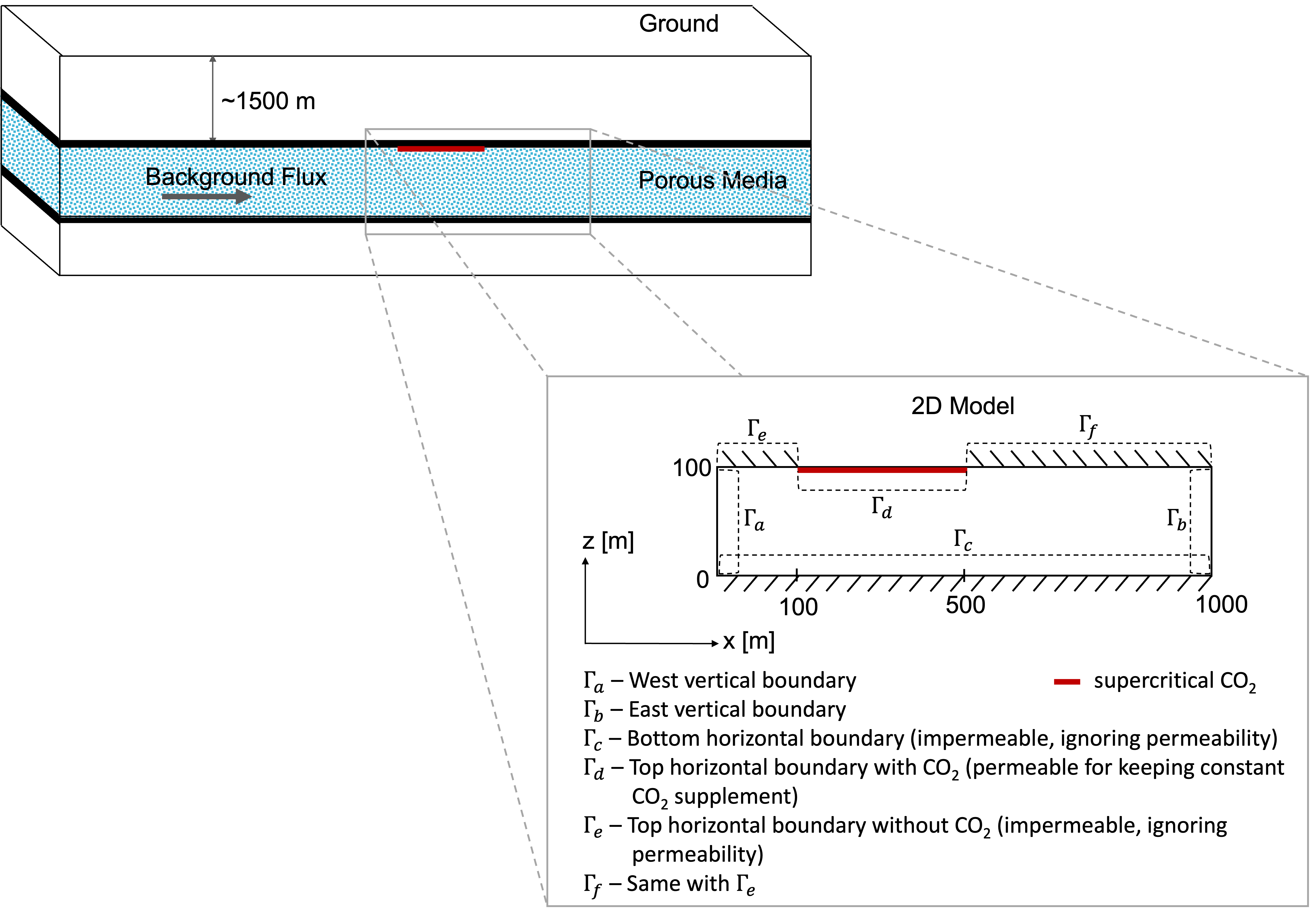}
		\caption{Diagram of the conceptual model, as translated into the two-dimensional simulation domain.}	
		\label{fig: parametric_study_4_model_structure}
	\end{figure}

	These equations are solved subject to impermeable boundary conditions at the top and bottom of the domain, and open-to-flow, specified-pressure boundary conditions at the upgradient and downgradient boundaries of the system. In addition, part of the upper boundary of the domain (imagined adjacent to supercritical CO\ts{2} lens) is set to the saturation concentration. See Figure \ref{fig: parametric_study_4_model_structure} for the geometrical configuration employed.

	The PFLOTRAN MPHASE module \citep{Lichtner2015} was utilized that solves Equations 1-3. In addition, the following boundary conditions are employed in PFLOTRAN (denoted by letters as in Figure \ref{fig: parametric_study_4_model_structure}):
	\begin{align}
		\bm {\nabla} c\cdot\bm{n} &= 0 && \Gamma_a\cup\Gamma_b\cup\Gamma_c\cup\Gamma_e\cup\Gamma_f, \label{eq: BC start}\\
		c &= c_\mathrm{sat} &&\Gamma_d,\\
		P &= P_{0,a} - \rho_0 g z && \Gamma_a,\\
		P &= P_{0,b} - \rho_0 g z && \Gamma_b,\\
		\bm {q}\cdot\bm{n} &= 0 && \Gamma_c\cup\Gamma_d\cup\Gamma_e\cup\Gamma_f. \label{eq: bc stop}
	\end{align}
	Here, $P_{0,a}$ is the pressure at $(x,z)=(0,0)$, and $P_{0,b}$ is the pressure at $(x,z)=(0,L)$, where $L$ is the length of the model domain, and $\bm{n}$ is the face normal. $\rho_0$ represents the density of CO\ts{2}-free brine. Initial condition $c=0$ obtains everywhere in the domain.

\subsection{Boussinesq approximation and dimensionless groups}
	For analytical purposes, we follow numerous authors in employing the Boussinesq approximations and neglect density gradients. We non-dimensionalize, using accented characters to represent dimensionless quantities: $\hat{c}\equiv c/c_\mathrm{sat}$, $\hat{x} \equiv x/H$, $\hat{z} \equiv z/H$, defining the operator $\hat{\bm \nabla} \equiv \left<\frac{\partial}{\partial \hat{x}},\frac{\partial}{\partial \hat{z}}\right>$. We diverge from \cite{Riaz2006} and \cite{Emami-Meybodi2015a} by computing dimensionless time without reference to permeability, so that we may directly compare results with different permeabilities:
	\begin{equation}
		\hat{t}\equiv t\frac{D}{H^2}.
	\end{equation}
	It is convenient to introduce the quantity $\Delta_\rho$, representing the difference in density between CO\ts{2}-saturated and CO\ts{2}-free brine. It then follows that
	\begin{equation}
		\rho = \rho_0 + \Delta_\rho \hat{c}.
	\end{equation}
	Next, we non-dimensionalize pressure:
	\begin{equation}
		\hat{P} \equiv \frac{P + \rho_0 g z}{\Delta_\rho g H}.
	\end{equation}
	This translates into dimensionless boundary conditions:
	\begin{align}
		\hat{P} &= \frac{P_{0,a}}{\Delta_\rho g H} && \Gamma_a,\\
		\hat{P} &= \frac{P_{0,b}}{\Delta_\rho g H} && \Gamma_b.
	\end{align}
	It is convenient to re-express the boundary conditions directly in terms of background Darcy flux, $Q$. From Darcy's law
	\begin{equation}
		Q = \left<k\right>\frac{P_{0,a}-P_{0,b}}{\rho_0 g L} = \left[\frac{H}{L}\frac{\Delta_\rho}{\rho_0}\right]\left(\hat{P}\vert_a-\hat{P}\vert_b\right),
		\label{eq: Q def}
	\end{equation} 
	where $\left<\cdot\right>$ represents a volume averaged quantity. Instead of treating boundaries $\Gamma_a$ and $\Gamma_b$ as having a hydrostatic pressure distribution, we treat them as having uniform constant flux equal to that computed by applying Darcy's law across the domain as a whole. This is not an exact translation from PFLOTRAN, but it simplifies analysis and is equally physically defensible. We then define $\hat{\bm q} \equiv \bm{q}/Q$, and the dimensionless flow boundary conditions then become 
	\begin{align}
		\bm {q}\cdot\bm{n} &= -Q && \Gamma_a,\\
		\bm {q}\cdot\bm{n} &= Q && \Gamma_b.\\
	\end{align}
	Next, we introduce two dimensionless groups:
	\begin{eqnarray}
		\mathrm{Pe} &\equiv& \frac{Q H}{\theta D}, \label{eq: peclet}\\
		W &\equiv& \frac{k\Delta_\rho g}{\mu Q} \label{eq: W},
	\end{eqnarray}
	where Pe is a P\'{e}clet number, and $W$ is the ratio of buoyant (free convective) Darcy flux to mean background Darcy flux. Refactoring in terms of these dimensionless groups leads to a simplified system
	\begin{eqnarray}
		\frac{\partial \hat{c}}{\partial \hat{t}}&=&\hat{\bm {\nabla}}^2\hat{c}-(\mathrm{Pe})\hat{\bm {q}}\cdot\hat{\bm{\nabla}}\hat{c},\label{eq: dless conc}\\
		\hat{\bm {q}} &=& W (\hat{\bm{\nabla}}\hat{P}+\hat{c}\bm{k}),\label{eq: dless flux}\\
		\hat{\bm\nabla}\cdot\hat{\bm q}&=&0 \label{eq: dless conservation}.	
	\end{eqnarray}
	Here, $\bm k$ is the unit vector in the positive $z$ direction. The boundary conditions for the non-dimensional system are 
	\begin{align}
		\hat{c} &= 0 && \Gamma_a\cup\Gamma_b,\\
		\hat{c} &= 1 && \Gamma_d,\\
		\hat{\bm \nabla}\hat{c}\cdot\bm{n} &= 0 && \Gamma_c\cup\Gamma_e\cup\Gamma_f,\\
		\hat{\bm {q}}\cdot\bm{n} &= 1 && \Gamma_a\cup\Gamma_b,\\
		\hat{\bm {q}}\cdot\bm{n} &= 0 && \Gamma_c\cup\Gamma_d\cup\Gamma_e\cup\Gamma_f.
	\end{align}
	Thus, if the parameters describing (\ref{eq: peclet}-\ref{eq: W}) are spatially homogeneous, the entire system has two degrees of freedom, whose quasi-steady plume and uptake dynamics are represented by some, generally nonlinear function of these two parameters. We observe that
	\begin{equation} 
		\mathrm{Ra} \equiv \frac{k \Delta \rho g H}{\mu \theta D} = W\ \mathrm{Pe}. 
	\end{equation}
	where Ra represents the system Rayleigh number, and that if \eqref{eq: dless flux} is substituted into \eqref{eq: dless conc} and \eqref{eq: dless conservation}, yields a pair of equations in $\hat{P}$ and $\hat{c}$, containing only Ra (as $\hat{\bm \nabla} W=\bm{0}$ under homogeneous conditions). This may be a more familiar formulation. Here, the reformulation of the boundary conditions in terms of $\hat{P}$ re-introduces dependency on $W$, accounting for the second degree of freedom. For a convenient comparison with previous works, we will express our numerical results in terms of Ra and Pe. 
	
	A dimensionless mass uptake rate is developed as follows: where $\dot{M}$ is the dimensional molar uptake rate for the whole system, and $A$ is the area of the supercritical CO\ts{2} pool interface with the aquifer, it follows that the average flux at the boundary is $\frac{\dot{M}}{A\theta}$. Because immediately adjacent to the pool, flux is diffusive only,
	\begin{equation}
		\frac{\dot{M}}{A \theta}=D\left<\bm{\nabla} c\cdot \bm{n}\right>_{\Gamma_c}=\frac{D c_\mathrm{sat}}{H}\left<\bm{\hat{\nabla}} \hat{c}\cdot \bm{n}\right>_{\Gamma_c},
	\end{equation}
	where $\left<\cdot\right>_{\Gamma_c}$ indicates the average value over boundary $\Gamma_c$. As established, under homogeneous conditions where a quasi-steady plume develops immediately below the pool, $\bm{\hat{\nabla}} \hat{c}$ is determined exclusively by Pe and Ra. Thus, it is natural to employ the dimensionless uptake rate (Sherwood number):
	\begin{equation}
		\rm{Sh} \equiv \frac{\dot{M}H}{A\theta D c_\mathrm{sat}}.
		\label{eq: sherwood}
	\end{equation}
	
	To analyze heterogeneous systems, we define the following averaged quantities and seek to uncover the inter-relationship between them and dimensionless heterogeneity statistics:
	\begin{eqnarray}
		\tilde{\rm Pe} &\equiv& \frac{Q H}{\left<\theta\right> D}, \label{eq: avg peclet}\\
		\tilde{\rm Ra} &\equiv& \frac{\left<k\right>\Delta_\rho g H}{\mu \left<\theta\right> D} \label{eq: avc rayleigh},\\
		\tilde{\rm Sh} &\equiv& \frac{\dot{M}H}{A\left<\theta\right> D c_\mathrm{sat}}. \label{eq: avg sherwood}
	\end{eqnarray}

\subsection{Permeability heterogeneity and disruption of convective fingers}
	According to \cite{Szulczewski2013}, the finger wavelength at the onset of convection, $\lambda_c$, may be approximated as $\lambda_c = 90 D / V$, where $D$ is the Fickian diffusion constant, and $V$ is the maximum vertical velocity due to convection. The same authors also report the time for full finger formation, $t_f$ to be \begin{equation} t_f \approx 2000 \left(\frac{D}{V^2}\right). \end{equation} From the macrodispersion theory, we expect the horizontal spatial variance, $\sigma^2$, of solute that begins in a vertical line (i.e., a finger) to satisfy $\sigma^2(t) = 2 D_\infty t$. From a triangular approximation to the Gaussian, we expect the fingering pattern to be eradicated when $\lambda_c = 2 \sigma$. Define the time of eradication to be $t_e$. Then \begin{equation} t_e = \left(1012.5 \frac{D}{D_\infty}\right)\left(\frac{D}{V^2}\right). \end{equation} Since in general $D_\infty \gg D$, we expect that $t_f \gg t_e$ (i.e. the time of eradication will be small relative to the time of formation). As the growth and eradication processes are acting simultaneously, scaling considerations indicate the unlikelihood of finger formation under heterogeneous background flow; numerical computations are required to substantiate these conclusions.

\section{Simulations} 

\subsection{Characteristics of geological formation and fluid} 
	We selected parameter values characteristic of deep saline sandstone aquifer CCS projects, in line with site values summarized in \ref{tab: literature data}. For all our simulations, pressure at the datum $(x,z)=(0,0)$ was set to $\rm 20\ MPa$, and temperature assumed to be everywhere $\rm \ang{50}\ C$. These conditions are well beyond the critical point for CO\textsubscript{2} at $\rm 7.39\ MPa$ and \SI{31.1}{\celsius} \citep{Benson2008,Mathieu2006}, and local pressure fluctuations never reduced solubility to the point that a separate supercritical phase formed withing the domain. Consequently, geological conditions were of single-phase density-driven flow.
	
	Permeability fields were generated with geometric mean permeability, $\left<k\right>$, ranging from $\rm 10^{-14}\ m^2$ to $\rm 10^{-12}\ m^2$, assumed to be locally isotropic. Where heterogeneous permeability was employed, we assumed that log-permeability was normally distributed, with variance, $\sigma^2_{\ln k}$ ranging from 0 to 2, and horizontal ($x$-direction) correlation length ranging from 10 to 100 m. In all cases a 10:1 ratio of horizontal to vertical correlation length was maintained. Permeability and porosity are typically positively correlated. Preliminary simulations indicated use of uniform mean porosity with heterogeneous permeability yielded unacceptably large (30\%+) errors in CO\ts{2} uptake versus when suitably correlated porosity fields were employed. In all cases, local porosity was derived frm local permeability via the empirical Kozeny-Carman equation \citep{pape2000variation}:
	\begin{equation}
		k = \frac{\theta ^3}{2.5(1-\theta )^2 S^2}.
		\label{eq: K-C}
	\end{equation}
	Here, $S$ is the average specific surface area for quartz grain, computed as $S=6/d$ where $d=111.75\ \mathrm{\mu m}$ \citep{bear1972dynamics, Beckingham2017}. Even when spatially heterogeneous, we porosity and permeability changes due to pressure fluctuations were assumed negligible \citep{Al-Yaseri2017}, so that these parameters remained constant at each location throughout the simulation.
	 
	Dissolved CO\ts{2} concentration was assumed to be always at saturation adjacent to the supercritical source lens. Saturation concentration is a function of temperature and pressure \citep{Kolenkovic2014}, as well as salinity. Based on literature reviewed in Table \ref{tab: literature data} we assumed 10\% salinity in all simulations. Using the thermodynamic model of \cite{Duan2003}, we derived maximum solubility $\rm 0.9382\ mol\ kg^{-1}$ at \SI{50}{\celsius} and $\rm 20\ MPa$. This corresponds to a CO\ts{2} mole fraction $0.0166$ for use with PFLOTRAN.

%
%
%
%
%
%

	Background flow velocities in deep saline aquifers are generally small, less than $\rm 5\ m\ y^{-1}$, and may be arbitrarily small. Relevant literature velocities include \cite{Suckow2020}, who estimated a velocity of $\rm 1.2\ m\ y^{-1}$ from isotopic considerations in a case study aquifer, with higher velocities shown by \cite{patterson2005cosmogenic} at kilometer depths in the Nubian aquifer system. Darcy flux measurements of $\rm 0.33\ m\ y^{-1}$ by \cite{Wei1990} and $\rm 0.39\ m\ y^{-1}$ were reported at supercritical depths by \cite{hortle2009hydrodynamic} in respective case studies, and similar maximum flux was employed by \cite{Heinemann2016} in a detailed numerical parametric study of CO\ts{2} dissolution. The range of background fluxes examined was selected to reflect similar velocities. The target flux, $Q$, was indirectly specified by adjustment of the horizontal hydraulic gradient and Darcy's law via \eqref{eq: Q def}.

\subsection{Computational considerations}
\label{model_setup}

	All simulations employed the geometry shown in Figure \ref{fig: parametric_study_4_model_structure}, and obeyed the boundary conditions (\ref{eq: BC start}-\ref{eq: bc stop}). A functionally 2D geometry was implemented in the 3D PFLOTRAN with a structured grid of dimension $\rm 1000\times 1\times 100\ m^3$, where each grid block had dimensions $\rm 1\times 1 \times 1\ m^3$.

	Numerical simulations considered the effects of four meta-parameters: mean permeability, $\left<k\right>$, permeability log-variance, $\sigma_\mathrm{\ln k}^2$, horizontal correlation length, $\lambda _x$, and mean background Darcy flux magnitude, $Q$. For heterogeneous simulations ($\sigma_\mathrm{\ln k} ^2 > 0$), the \texttt{GSTools} \citep{gmd-15-3161-2022} library was used to generate spatial random permeability fields with exponential variogram, and manifesting the correct permeability mean, log-variance, and spatial correlation structure. These fields were discretized so that each PFLOTRAN grid block was assigned a constant value. The porosity for each grid block was determined from its corresponding permeability via \eqref{eq: K-C}.
	
	Hundreds of simulations were run (334 in heterogeneous domains, 28 in homogeneous domains), covering the intersections of all four meta-parameters, with multiple random fields (typically five) employed for heterogeneous domains. Homogeneous simulations covered three domain permeabilities ($\rm 10^{-14}-10^{-12}\ m^2$) and a range of design background Darcy fluxes covering the range $\rm 0-5\ m\ y^{-1}$. Heterogeneous simulations covered the same range of permeabilities and design Darcy fluxes, but also considered three log-variances $\sigma^2_\mathrm{\ln k}=0.5$, $1$, and $2$, and two horizontal correlation lengths: $10$ and $\rm 100\ m$.

	The total number of moles that have dissolved into the aquifer at time $t_e$ since the beginning of each simulation is computed by integrating molar concentration over the entire pore space of the domain, and integrating the advective molar fluxes  over the upgradient and downgradient boundaries---boundaries $\Gamma_a$ and $\Gamma_b$---and over time (the concentration boundary conditions ensure all fluxes are advective at those locations):
	\begin{equation}
	 M_{CO_2}(t_e)=W\left[ \int^L_0 \int^H_0 c\theta \vert_{t=t_e} dz dx + \int^H_0 \int^{t_e}_1 c \bm{q}\cdot\bm{n}\vert_{x=0} - c \bm{q}\cdot\bm{n}\vert_{x=L} dt dz\right].
		\label{CO2-mass} 
	\end{equation}
	We observed (see for example Figures \ref{8.3.41-hete-short-length} and \ref{8.3.42-hete-long-length}) that the  of CO\ts{2} dissolution within particular realizations were approximately constant over time, so it is reasonable to compute a single molar uptake rate:
	\begin{equation} \centering \dot{M}_{CO_2} \approx \frac{M_{CO_2}(t_\mathrm{max})}{t_\mathrm{max}}, \label{CO2-rate} \end{equation}
	where $t_\mathrm{max}$ is the maximum time of the simulation.

%

\section{Results and discussion} 
	
\begin{figure}
		\centering \includegraphics[width=\linewidth]{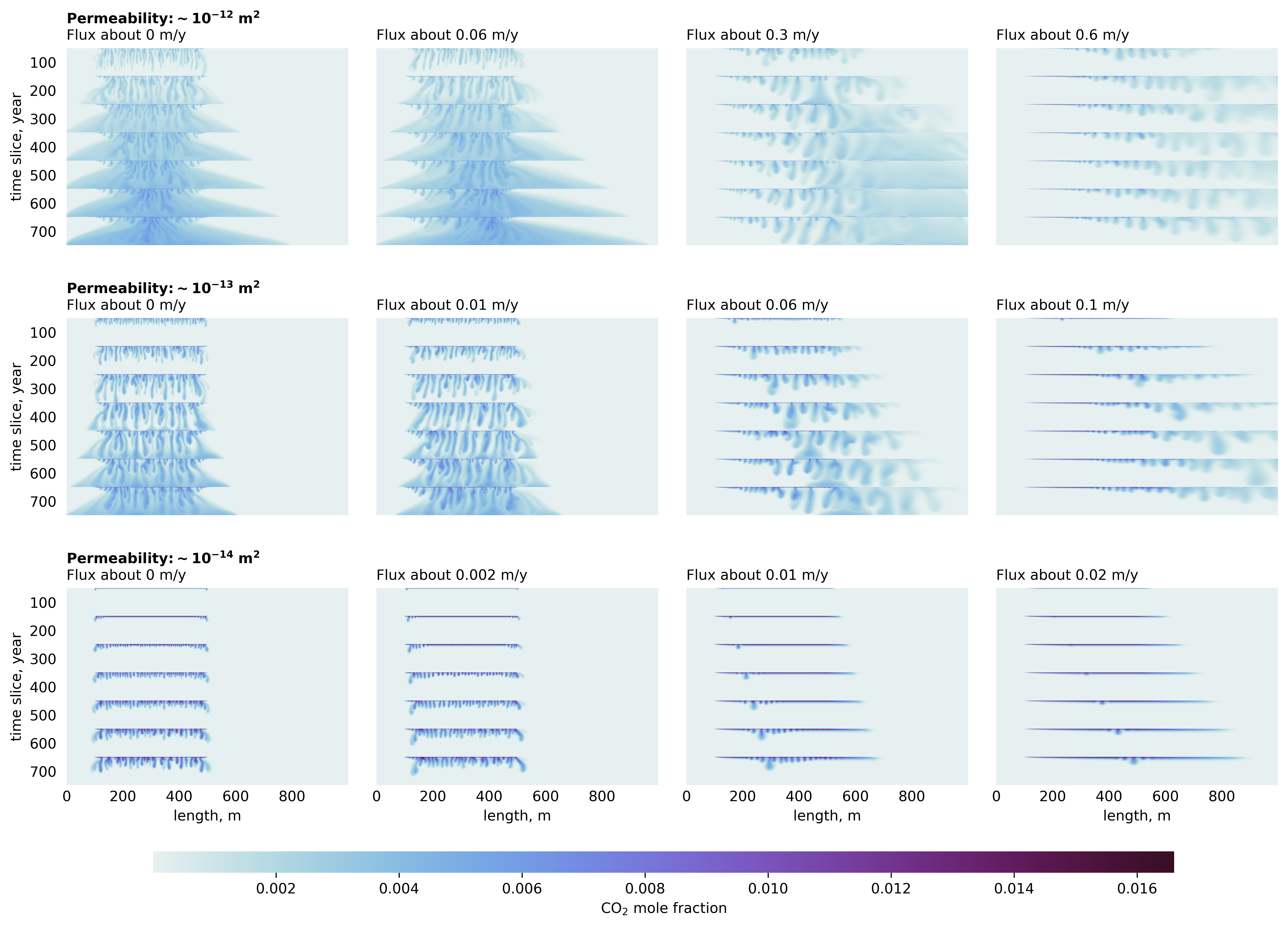} 
		\caption{Time series color maps showing CO\textsubscript{2} mole fraction (darker color indicates higher concentration) for a variety of background fluxes and spatially homogeneous aquifer permeabilities. Each time series shows plume evolution spanning $\rm 100$ y to $\rm 700$ y after introduction of supercritical CO\textsubscript{2}} \label{parametric_study_1_homogeneity_process} 
	\end{figure}

	Figure \ref{parametric_study_1_homogeneity_process} illustrates the CO\textsubscript{2} migration at every 100 years during $\rm 700$ years in homogeneous aquifers with three different permeabilities $\rm k$ ($\rm log_{10}k=-12,-13,-14$), each with four different background Darcy fluxes. The effect of increased permeability is apparent in increasing the rate of gravitational convection, and also in generating more diffuse, less pronounced fingers. The effect of increased background flux is apparent in generating fewer, more widely spaced fingers, and ultimately suppressing them altogether. Interestingly, for even modest background fluxes, the convection enters a transitional regime in which it ceases to interact with the bottom of the aquifer: here the classical sequence of fingering followed by shutdown and slumping, commonly seen in numerical simulations without background flow \citep[e.g.,][]{Bolster2014a}, never manifests.

\begin{figure}[h] 
		\centering \includegraphics[width=\linewidth]{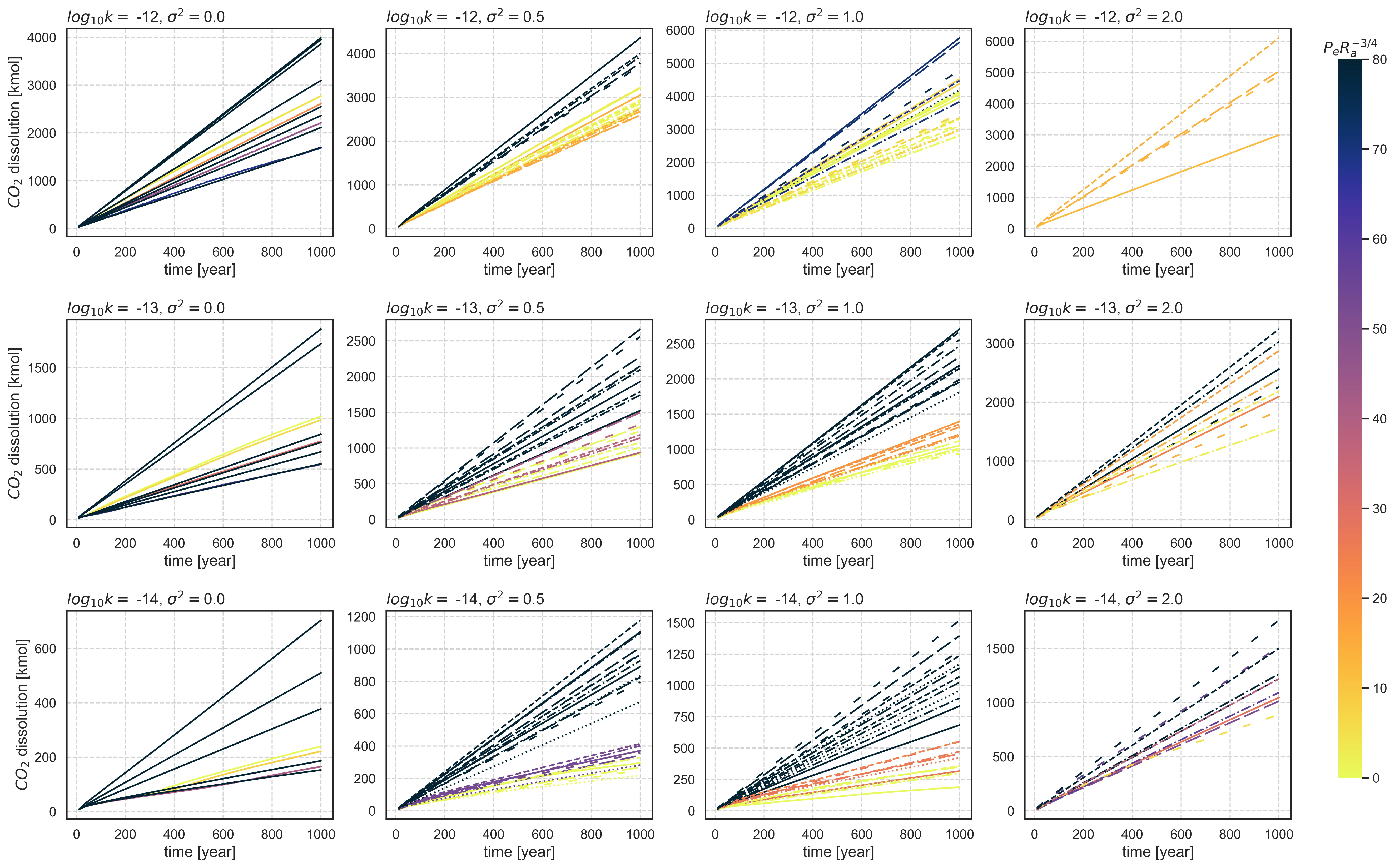} \caption{Cumulative CO\textsubscript{2} mass is stored in the aquifer for 1000 years at different heterogeneous permeability fields with variance (0, 0.5, 1.0, 2.0) and short horizontal correlation length 10 m. Four background fluxes with the same permeability and variance are shown. Lines are colored according to $\rm Pe Ra^-{3/4}$. Within each axis, each line of the same style represents a simulation in the same permeability field realization.} \label{8.3.41-hete-short-length} 
	\end{figure}

\begin{figure}[h] 
		\centering 
		\includegraphics[width=\linewidth]{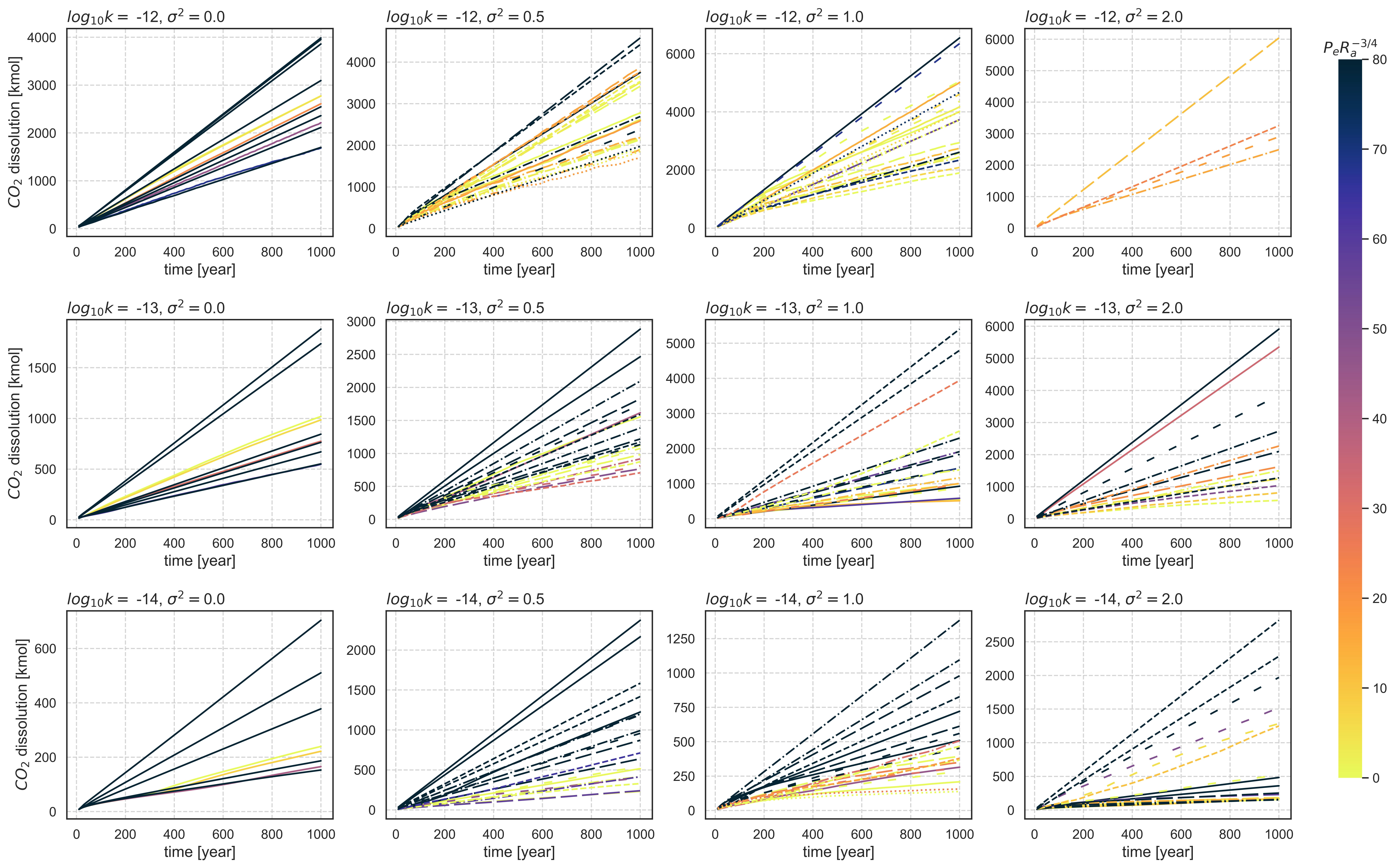} \caption{Cumulative CO\textsubscript{2} mass has been stored in the aquifer for 1000 years at different heterogeneous permeability fields with variance (0, 0.5, 1.0, 2.0) and long horizontal correlation length 100 m. Four background fluxes with the same permeability and variance are shown. Lines are colored according to $\rm Pe Ra^-{3/4}$. Within each axis, each line of the same style represents a simulation in the same permeability field realization.} \label{8.3.42-hete-long-length}
	\end{figure}

	For heterogeneous domains, cumulative uptake over time computed with \eqref{CO2-mass} is shown, grouped by log-variance, mean permeability, and background Darcy flux. Results for $\rm 10\ m$ horizontal correlation length are shown in Figure \ref{8.3.41-hete-short-length}, and for $\rm 100\ m$ horizontal correlation length in Figure \ref{8.3.42-hete-long-length}. Several patterns are apparent. In general, there is significant inter-realization variability for realizations sharing the same meta-parameters, though individual realizations all display near-linear uptake over time. Within each of the axes in these figures, a particular line style indicates a particular permeability field realization: these were reused at different background flow rates. It is apparent that particular realizations are associated with consistently elevated or depressed CO\ts{2} dissolution rates due to the specific configuration of their porosity-permeability field. 
	
	Inter-realization variability is seen to be generally associated with greater permeability log-variance, and also with greater horizontal correlation length, possibly due to the length scale of low permeability features increasing relative to other system length scales such as aquifer depth CO\ts{2} pool length. As in the homogeneous simulations, higher permeability is associated with faster uptake. However, the relationship with background flux is less obvious and requires quantitative analysis.
	
\begin{figure}[!ht]
		\centering
		\includegraphics[trim={0 5cm 0 0}, clip, width=\linewidth]{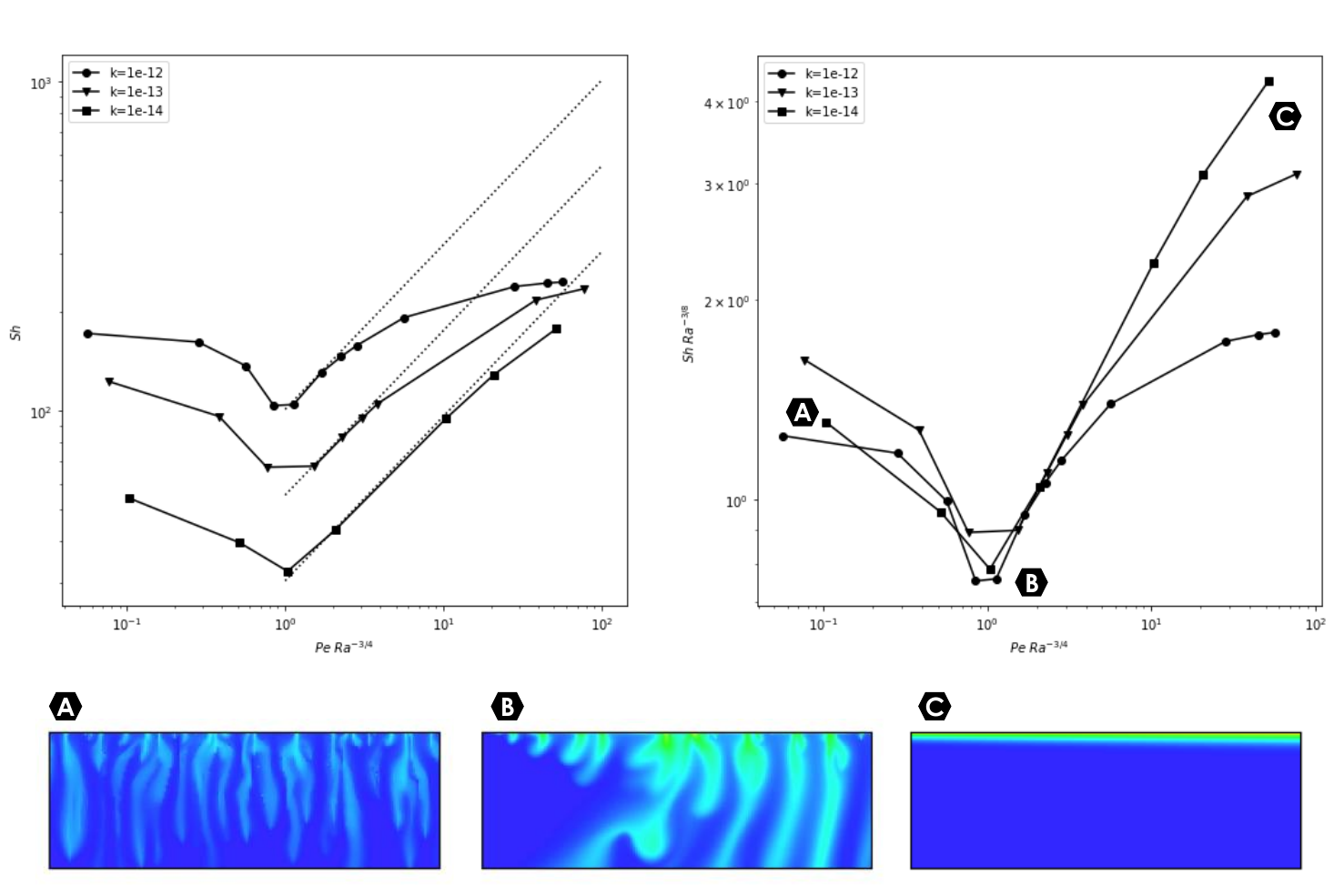} 
		\caption{Dimensionless uptake rate as a function of Pe and Ra for simulations in homogeneous domains featuring three different permeabilities. Curves are annotated with color maps of CO\textsubscript{2} concentration beneath the source extracted from high-permeability simulation after $\rm 200\ y$. At small background velocities (small $\rm Pe\ Ra^{-3/4}$), gravitational fingers are apparent. At intermediate velocities, mixed convection is apparent, and the uptake rate is minimized. At large background velocities, fingers are absent (the CO\textsubscript{2} distribution resembles an LNAPL dissolution plume), and uptake is maximized.}
		\label{fig: transition}
	\end{figure}

	For each simulation, we computed the quasi-constant uptake rate with \eqref{CO2-rate} and converted it to a non-dimensional Sherwood number with \eqref{eq: sherwood}. Our earlier analysis suggests that, for homogeneous domains, a fundamental relationship may be found linking  Sh, Ra, and Pe, exclusively. We seek to identify it empirically by selecting quantities that unify the uptake results obtained using various permeabilities on log-log axes. Beginning with a Sh vs. Ra plot, we identified the dimensionless groups  $\rm Pe\ Ra^{-3/4}$ and $\rm Sh\ Ra^{-3/8}$. These are plotted against each other for each of the three geometric mean permeabilities in Figure \ref{fig: transition}. Minimum uptake was identified consistently with the point $\rm Sh\ Ra^{-3/4} = 1$. By inspection of plume concentration below the CO\ts{2} pool, this was identified with disruption of CO\ts{2} fingers during the transition from gravitation-dominated free convection regime at low background fluxes to advection-dominated forced convection corresponding to high background fluxes. Examples are shown in the figure. Analytical work by \cite{Emami-Meybodi2015a}, assuming negligible gravitational convection with high background fluxes, suggested a relationship of the form $\rm Sh \propto Pe^{1/2}$ for higher background fluxes. We note that the qualitative behavior is congruent with that found by \cite{Emami-Meybodi2015a}, as well as \cite{Tsinober2022}, who considered only a single permeability, but the dimensionless scaling relationship is original. The same scaling is recovered against our dimensionless quantities in the lower-velocity regions of the advection-dominated regime, again shown in the same figure. We find then that at moderately high velocities in homogeneous media
	\begin{eqnarray}
		\rm Sh\ Ra^{-3/8} &\propto& \left(\rm Pe\ Ra^{-3/4}\right)^{1/2},\\
		\rm Sh &\propto& \rm Pe^{1/2},
	\end{eqnarray}
	corroborating our selection of dimensionless groups against the analytic solution and recovering the permeability-independence expected in this regime. We note that this relationship breaks down at very high background fluxes. We posit that this is due to PFLOTRAN MPHASE using the exact equations instead of the Boussinesq approximations. In PFLOTRAN's formulation, there is an additional diffusion-like flux $\frac{k g z}{\mu} \nabla \rho \propto \frac{k g z}{\mu} \nabla c$ resulting from \eqref{eq: PFLOTRAN flux} which competes with the diffusive flux $\theta D \nabla c$ in \eqref{eq: PFLOTRAN mass balance}. As the value of its coefficient increases, this increasingly dominates, and invalidates the assumption of permeability-independence underlying the analytical solutions in \cite{Emami-Meybodi2015a}. This supposition is corroborated by the observation from Figure \ref{fig: transition}, that there is an increased deviation from the $\rm Pe ^{-1/2}$ trend line with increased permeability.
	
	For heterogeneous simulations, a similar theory-guided empirical approach was adopted. Dimensionless uptake rates for the various simulations were initially plotted, employing the dimensionless groups identified in the homogeneous analysis, except the averaged groups $\rm \tilde{Sh}$, $\rm \tilde{Pe}$, and $\rm \tilde{Re}$ were used. Points were differentiated by mean permeability, permeability log-variance, horizontal permeability correlation length, and dimensionless groups on the axes. The plotted dimensionless groups were modified empirically to (a) remove additional trends by point category (e.g., log-variance) relative to the ensemble trend while (b) containing the groups identified in the homogeneous analysis as a special case. Correlation length was not seen to affect average performance, and the dimensionless groups identified were  $\tilde{\rm Pe}\ \tilde{ \rm Ra}^{-3/4} (1 + \sigma^4_\mathrm{\ln k})$ and $\tilde{\rm Sh}\ \tilde{ \rm Ra}^{-3/8} (1 + \sigma^4_\mathrm{\ln k})$. Again consistent with the homogeneous analyses of \cite{Emami-Meybodi2015a}, a horizontal trend line was found to be supported in the low background flux (gravitation dominated) regime, and the previously-identified slope $1/2$ log-log trend line was found to be supported in the high background flux regime. From these relationships, the following scaling relations obtained in the gravitational convection-dominated (low background flux) regime:
	\begin{align}
		\tilde{\rm Sh} &\propto \tilde{\rm Ra}^{3/8} & \sigma^2_\mathrm{\ln k} \ll 1,\\
		\tilde{\rm Sh} &\propto \frac{\tilde{\rm Ra}^{3/8}}{\sigma^4_\mathrm{\ln k}} & \sigma^2_\mathrm{\ln k} \gg 1.
	\end{align}
	In the advection-dominated (high background flux) regime:
	\begin{align}
		\tilde{\rm Sh} &\propto \tilde{\rm Pe}^{1/2} & \sigma^2_\mathrm{\ln k} \ll 1,\\
		\tilde{\rm Sh} &\propto \frac{\tilde{\rm Pe}^{1/2}}{\sigma^2_\mathrm{\ln k}} & \sigma^2_\mathrm{\ln k} \gg 1.
	\end{align} 
	Across all regimes, heterogeneity was found to be associated with lower expected uptake. Strikingly, even for the relatively modest heterogeneities considered here, profound inter-realization variability limited the predictive value of the dimensionless scaling relationships. 
%

	\begin{figure}[h] 
	\centering 
	\includegraphics[trim={0 7cm 0 0}, clip, width=\linewidth]{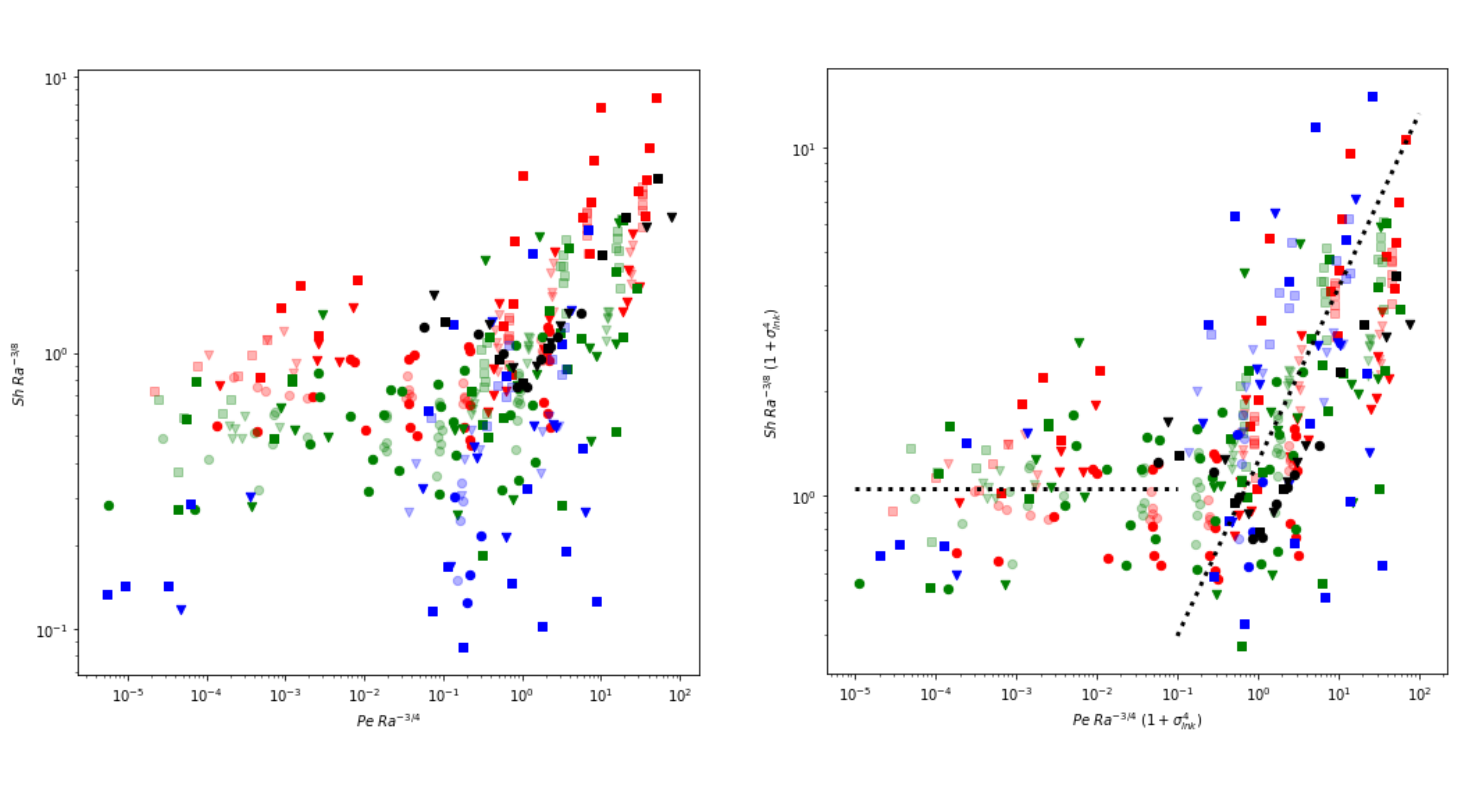} 
	\caption{Dimensionless uptake rates versus Rayleigh number and Peclet number respectively. Points are color-coded according to $\sigma^2_\mathrm{\ln k}$ (black: 0, red: 0.5, green: 1, blue: 2), with shape representing mean log-permeability (circle: $10^{-12}$, triangle: $10^{-13}$, square: $10^{-14}$), and saturation representing horizontal correlation length (saturated: $100 \rm{m}$, desaturated: $10 \rm{m}$). At left, points are shown in terms of the dimensionless groups developed for the homogeneous analysis. At right, points are shown in terms of new empirical groupings that eliminate some residual trends not captured by the previous groupings.}
	\label{dimenionless-assemble} 
	\end{figure}

\section{Summary and conclusions} 

	We have presented a large-scale Monte Carlo study employing high-resolution numerical simulations of dissolution, gravity-driven CO\textsubscript{2} convection, and background flow in deep saline aquifers featuring multi-Gaussian permeability fields. Key scientific aims were to understand how subsurface heterogeneity and flow impact the CO\textsubscript{2} uptake rate, as well as the extent to which analyses based on various simplifying assumptions are applicable to understanding more realistic conditions. Our study considered the impact of four key parameters characterizing heterogeneous background flow: permeability (i) geometric mean, (ii) log-variance, and (iii) correlation length, as well as (iv) average horizontal background flow velocity. 

	A number of conclusions can be drawn from our analysis:
	\begin{itemize} 
		\item[\ding{227}] In homogeneous permeability fields, three uptake regimes are apparent, associated with increasing background flow: gravitation-dominated, transition, and advection-dominated, with uptake minimized in the transition regime. This is qualitatively consistent with prior studies \citep{Emami-Meybodi2015a,Tsinober2022}. Unification of data from multiple permeabilities suggested the novel criterion $\mathrm{Pe\ = Ra^{3/4}}$ for regime transition, and this was corroborated by heterogeneous permeability simulations.
		\item[\ding{227}] Background flow was found to significantly enhance CO\textsubscript{2} dissolution, with expected uptake scaling as $\mathrm{Pe}^{1/2}$ in the advection-dominated regime. This regime was associated with dimensional pore water velocities above $\rm 5\ m\ y^{-1}$ for permeability $10^{-12}\ \mathrm{m^2}$, and $\rm 0.5\ m\ y^{-1}$  for permeability $10^{-14}\ \mathrm{m^2}$, and so may not be realistic for many deep saline aquifers. 
		\item[\ding{227}] Mean permeability was found to play a dominant role in determining CO\ts{2} uptake rate in the gravitation-dominated regime, but not to be relevant in the advection-dominated regime.
		\item[\ding{227}] Permeability heterogeneity was found to display a qualitatively significant effect on fingering patterns and to quantitatively impact CO\textsubscript{2} uptake in all regimes. Expected uptake decreases proportionally with variance in the advection-dominated regime, and as the \textit{square} of variance in the gravitational convection-dominated regime. Increased correlation length was not found to significantly affect expected uptake, but did increase its variance.
		\item[\ding{227}] In general, permeability heterogeneity both decreased mean uptake rates and increased performance variance, even at the modest heterogeneity levels considered in this study and so is undesirable from the perspective of CCS scheme design. Though in all realizations, a quasi-steady uptake rate developed, profound inter-realization variability was apparent. This limits our ability to predict CO\ts{2} uptake performance \textit{a priori}.
		\item[\ding{227}] The homogeneous simulations display a reduced CO\ts{2} dissolution rate at high P\'{e}clet numbers relative to predictions from analytic solutions that neglect gravitational effects, particularly at higher permeabilities. This is contrary to what might be expected in the forced convection regime, and we attribute this effect to a transport term captured by PFLOTRAN that is neglected under the Boussinesq approximation. This approximation may not be harmless and may result in CO\ts{2} dissolution rates being overestimated at high background flow rates.
	\end{itemize}
	
	Our simulations point towards some directions for future research. Whereas previous studies have focused on diffusion only, the macrodispersive effect of heterogeneous horizontal flow on disrupting finger formation and influencing CO\textsubscript{2} dynamics was readily apparent. It was not possible to study this effect quantitatively with PFLOTRAN, which does not model dispersion in its MPHASE mode. Further, all Eulerian codes are affected by numerical dispersion. It would be enlightening to study the interaction of permeability heterogeneity, flow and dispersion with an alternative numerical scheme. Also interesting would be to analyze the effect of permeability heterogeneity on fingering patterns, as a strong effect was observed but not quantified. 

	\section*{Acknowledgments}
		SKH acknowledges the support of ISF personal research grant 1872/19 and holds the Helen Ungar Career Development Chair in Desert Hydrogeology.
		SK thanks Environmental Molecular Sciences Laboratory for its support. 
		Environmental Molecular Sciences Laboratory is a DOE Office of Science User Facility sponsored by the Biological and Environmental Research program under Contract No. DE-AC05-76RL01830.

\singlespacing 
\bibliographystyle{myplainnat} 
\bibliography{MS_CO2,co2_skh}

\pagebreak
\appendix

\begin{landscape} 

	\begin{table}
		\caption{Summary of literature values of geological and hydraulic properties relevant to CO\ts{2} dissolution in characteristic deep aquifers.}
		\renewcommand\arraystretch{1.5}
		\resizebox{\linewidth}{!}{
			\begin{tabular}{llccccccccl}
				Reference 	& Stratum	& Depth [m] & Thickness [m]	& Pressure [MPa] & Temperature [$\SI{}{\celsius}$] & Permeability [$\rm m^2$] 			& Porosity [\%]		& Salinity [\%]  & Velocity [$\rm m\ y^{-1}$] 	& Notes \\  
				\hline
				\cite{Kumar2020} 						& Dollar Bay Formation, USA 				&	$\sim 2750$				&	$\sim 70$				&  $\sim 39$ 				& $\sim 75$ 		& $\sim 1-100 \times 10^{-15}$		&	$\sim 1.5-19$				&  $\sim 30$					 & -	& \\
				\cite{doughty2008site,sakurai2006monitoring}  
				&  Frio Formation, USA &	$\sim 1500$				&	$\sim 23$		& $\sim 15.2$ 		& $\sim 56-65$ 				      & $\sim 1\times 10^{-12} $  		&	$\sim 28$				& $9.3$  							&  - & Case study	\\
				\cite{Bachu1994,bachu1993hydrogeology}  
				& Alberta basin, Canada
				& $\sim 800 - 2000$		& $\sim 200$	& 	$> 8$ 				& $\sim 30$	 					  & $ \sim 1-15 \times 10^{-15}$	& $\sim 10-12$		& 		$\sim 5.5-35$				& $\sim 0.01-0.1$ 											& CCS project\\
				\cite{spence2014peterhead,shell2016} 	& Goldeneye Field, UK		&	$> 2000$			&	-		&  $\sim 26$  			& $\sim 83$  			& $\sim 6.9-15\times 10^{-13}$  	&	$\sim 25$			&  	$5.4$								& -  				& CCS project \\
				\cite{kikuta2005field,xue2006estimation}& Haizume Formation, Japan &	$\sim 1100-1300$	&	$\sim 60$		&  $\sim 11$ 	 	& $\sim 48$  				      & $\sim 7\times 10^{-15}$  		&	$\sim 23$		&  - & - 										& Pilot-scale injection site \\			
				\cite{streibel2014pilot,akono2020influence}
				& Illinois Basin, USA &		$\sim 1691 - 2150$		&	-		& $\sim 23$ 				& $\sim 60$ 		& $\sim 2\times 10^{-13}$ 			&	$\sim 20$		&  		$ \sim 12$							&  - 				& CCS project (IBDP)	\\ 
				\cite{Whittaker2004}		 			& Midale Aquifer, Canada	& $\sim 1400$			&	$\sim 23$ 	&	$\sim 15$			& 	$\sim 63$					  & $\sim 3.5\times 10^{-14}$ 		&	$\sim 13$		& 	$\sim 5-15$						& 					& CCS project	\\
				\cite{patterson2005cosmogenic,Robinson2007} 
				& Nubian Aquifer, Egypt &	$> 800$					&	$\sim 300-600$	& 	$\sim 10$		& 	$ \geq 26 $					  & 	$ \leq 5\times 10^{-12}$ 	&	$\sim 20$		& 			$ < 0.04$ 				& $ \sim 0.5-3.5$  				& 	 \\
				\cite{Jiang2014a,luo2014ordos,zunsheng2014opportunity}  
				& Ordos Basin, China	&	$1680-2500$					&	$\sim 20-40$		& $\sim 8-25$  			& $\sim 56-69$  		& $\sim 3.5-40\times 10^{-16}$  	&		$\sim 10$		&   	$\sim 1-7$							& -  & CCS demonstration project	\\
				\cite{koperna2014project,conaway2016comparison} 
				& Paluxy Formation, USA &	$2880-2987$					&	$\sim 6-23$			& $\sim 28-30$  			& $\sim 20-50$  	& $\sim 1-400\times 10^{-14} $  	&	$\sim 13-24$			& 		$\sim 20$						    & -  & CCS project assessment	\\
				\cite{hansen2013snohvit,grude2014pressure}  
				& Tubåen Formation, Norway 	&	$ 2560 - 2670 $			&	$\sim 110$		&  $\sim 28-39$  			& $\sim 92-95$  	& $\sim 1-1000\times 10^{-14}$  	&		$\sim 7-20$		&  $\sim 14-16$ & - 									& CCS project	\\
				\cite{arts2008ten,chadwick2006geophysical,white2014geomechanical}		 
				& Utsira Formation, North Sea & $\sim 1000$			&	$>50$		& $\sim 8-11$ 		& $\sim 36-46$  				  	  &  	$ \sim 3 \times 10^{-12}$	&	$> 20$		& $ \sim 4$  						&		- 				& CCS project	 \\
				\cite{WURDEMANN2010938,PREVEDEL20146067}
				& Stuttgart Formation, Germany & $\sim 1000$ 	& 	$\sim 80$ 	& $\sim 8.2$ 		& 	$\sim 35$ 	& $\sim 5-100\times 10^{-14}$ 	& $\sim 5-35$ 							& - 					&	-  & CCS pilot project \\
				\cite{UNDERSCHULTZ2011922,hortle2009hydrodynamic}
				& Waarre Formation, Australia & $\sim 2000$		& $\sim 25$		& $19.59$			& 	-			& $\sim 7-10\times 10^{-13}$ & 	$\sim 13.8$					&	$\sim 2$		& $\sim 0.39$ 					&		CCS project \\
				\hline
			\end{tabular}
		}
		\label{tab: literature data}
	\end{table}

\end{landscape}

\end{document}